\documentclass[5p,twocolumn,times,numbers,sort& compress]{elsarticle}

\journal{Physics Letters B}

\usepackage{amsmath,amssymb}
\usepackage{graphicx}
\usepackage{bm}
\usepackage{hyperref}
\usepackage{tikz}
\usepackage{tabularx}
\usepackage{longtable}
\usepackage{xcolor}
\usepackage{csquotes}
\usepackage{dcolumn}
\usepackage{etoolbox}
\usepackage[T1]{fontenc}
\usepackage[utf8]{inputenc}

\biboptions{numbers,sort&compress}

\usepackage{etoolbox}

\makeatletter
\patchcmd{\thebibliography}
  {\sloppy\clubpenalty4000\widowpenalty4000}
  {\sloppy\clubpenalty4000\widowpenalty4000%
   \setlength{\itemsep}{0pt}%
   \setlength{\parsep}{0pt}%
   \setlength{\parskip}{0pt}%
   \setlength{\topsep}{0pt}%
   \linespread{1}\selectfont%
   \small}
  {}{}
\makeatother

\begin{document}

\title{Evidence of the Excited  X(5)-like Critical-Point Symmetry Structures in $^{152}$Sm}

\author[1,2]{S.~Basak\corref{cor2}}
\cortext[cor2]{Present address: Institute of Modern Physics, Chinese Academy of Sciences, Lanzhou, Gansu - 730 000, China}
\author[3]{S.~Rajbanshi}
\author[1,2]{T.~Bhattacharjee\corref{cor1}}
\cortext[cor1]{Corresponding author}
\ead{btumpa@vecc.gov.in}
\author[1,2]{D.~Kumar}
\author[1,2]{A.~Pal}
\author[4]{S.~S.~Alam}
\author[5]{A.~Saha}
\author[1,2]{A.~K.~Sikdar}
\author[1,2]{J.~Nandi}
\author[1]{Ananya~Das}
\author[6]{Shabir~Dar}
\author[7]{S.~Samanta}
\author[8]{S.~Chatterjee}
\author[8]{R.~Raut}
\author[8]{S.~S. Ghugre}
\author[9]{A.~Adhikari}
\author[10]{Y.~Sapkota}
\author[11]{R.~Rahaman}
\author[12]{A.~Gupta}
\author[11]{A.~Bisoi} 
\author[13]{S.~Sharma}
\author[12]{S.~Das}
\author[14]{P.~Das}
\author[14]{U.~Datta}
\author[15]{I.~Ray}
\author[16]{G.~Duch\^{e}ne}
\affiliation[1]{Variable Energy Cyclotron Centre, Kolkata - 700 064, India}
\affiliation[2]{Homi Bhabha National Institute, Training School Complex, Anushakti Nagar, Mumbai - 400 094, India}
\affiliation[3]{Presidency University, Kolkata 700073, India}
\affiliation[4]{Government General Degree College, Chapra - 741 123, West Bengal, India}
\affiliation[5]{ICFAI University Tripura, Agartala, Tripura-799 210, India}
\affiliation[6]{Department of Physics and Astronomy, Division of Applied Nuclear Physics, Uppsala University, 75120 Uppsala, Sweden}
\affiliation[7]{Adamas University, Kolkata, West Bengal 700126}
\affiliation[8]{UGC-DAE CSR, Kolkata Centre, Kolkata - 700 098, India}
\affiliation[9]{Ghani Khan Choudhury Institute of Engineering and Technology, Malda - 732141, India}
\affiliation[10]{Department of Physics, Dudhnoi College, Dudhnoi, Goalpara, Assam-783120, India}
\affiliation[11]{Indian Institute of Engineering Science and Technology, Howrah-711 103, India}
\affiliation[12]{Institute of Engineering and Management, University of Engineering and Management Kolkata, Kolkata-700160}
\affiliation[13]{Manipal University Jaipur, Jaipur-303007, Rajasthan, India}
\affiliation[14]{Saha Institute of Nuclear Physics, Kolkata-700 064, India}
\affiliation[15]{Jadavpur University, Kolkata-700 032, India}
\affiliation[16]{Universit\'{e} de Strasbourg, CNRS, IPHC UMR 7178, F-67 000 Strasbourg, France}

\begin{abstract}
The positive-parity structure of $^{152}$Sm has been investigated through high-statistics $\gamma$-ray spectroscopy following the ($^{150}$Nd($\alpha$,2n) $^{152}$Sm reaction at $E_{\mathrm{lab}}$ = 26 MeV. Several collective structures built on excited 0$^+$ states have been extended through the observation of new levels and $\gamma$-ray transitions, and spin-parity assignments have been established using directional-correlation and linear-polarization measurements. Electromagnetic transition strengths (B(E2)), deduced from measured branching ratios and known level lifetimes, reveal pronounced collectivity among the excited configurations. The resulting level scheme provides evidence for a sequence of excited collective bands extending beyond the well-known ground-state and first excited 0$^+$ structures. The excitation energies and transition strengths are examined within the framework of the X(5) critical-point description of the first-order U(5)-SU(3) shape-phase transition. In addition to the established X(5)-like features of the low-lying spectrum, the observed systematics of the higher-lying bands are found to be consistent with excited collective structures exhibiting X(5)-like characteristics. The results provide new constraints on the realization of critical-point behavior in finite nuclei and on the evolution of collectivity in the N=90 region.
\end{abstract}

\begin{keyword}
Nuclear Structure, Level Lifetime, Dynamical Symmetries, Interacting Boson Approximation, X(5) Symmetry.
    
\end{keyword}
\maketitle

Phase transitions in physical systems are characterized by rapid changes in an order parameter as a control parameter crosses a critical value. Such behavior is encountered across diverse areas of physics, ranging from condensed matter to ultracold atomic systems. A notable example is the superconducting phase transition, where the emergence of long-range coherence is associated with the breaking of gauge symmetry. The concept of universality, wherein systems with distinct microscopic properties exhibit similar behavior near criticality, provides a unifying framework for understanding these phenomena~\cite{rfcasten11,f2,f3}.

Atomic nuclei offer a unique setting for the study of quantum phase transitions in finite, self-bound many-body systems. In contrast to infinite systems, where phase transitions are sharply defined, nuclei exhibit a gradual structural evolution driven by changes in nucleon number and shell occupancy. As a result, observables such as excitation energies, electromagnetic transition strengths $B(E2)$, and systematic trends in level structures serve as key signatures of underlying phase transitional behavior~\cite{rfcasten11,rfcasten22}. These features provide important insight into the emergence of collective motion and symmetry in strongly interacting finite systems. Within the framework of the Interacting Boson Approximation (IBA), nuclear structure is described in terms of the $s$- and $d$-bosons, giving rise to three dynamical symmetries: U(5) (spherical vibrator), SU(3) (axially deformed rotor), and O(6) ($\gamma$-soft rotor)~\cite{gsharff, abohr1953, lwilets}. Each limit is associated with characteristic spectroscopic signatures, such as the energy ratio $R_{4/2} = E(4^+_1)/E(2^+_1)$, which takes the values 2.00, 3.33, and 2.50 for U(5), SU(3), and O(6), respectively~\cite{fiachel1}. Nuclei located in transitional regions between these limits exhibit intermediate structural properties, requiring detailed spectroscopic studies to identify the nature of the underlying phase evolution.

Nuclei positioned at the critical point of a quantum shape-phase transition are particularly challenging to describe because of the coexistence and mixing of competing collective degrees of freedom. Analytic solutions of the Bohr Hamiltonian for specific potentials have led to the concept of critical-point symmetries (CPS), notably X(5) for first-order ~\cite{fiachel3}  and E(5) for second-order phase transitions ~\cite{fiachel2}, which describe nuclei at the critical points of shape phase transitions. The X(5) symmetry corresponds to the transition between the U(5) and SU(3), while E(5) describes the transition from U(5) and O(6). These solutions provide parameter-independent predictions (up to an overall scale) for excitation energies and transition rates. Within the X(5) critical-point symmetry, the excitation energies are denoted by $E_{s,L}$, where $s$ labels the successive zeros of the Bessel function and $L$ is the angular momentum quantum number $(L=0,2,4,\ldots)$. Despite their simplicity, these models provide a remarkably successful description of transitional nuclei. Although X(5) and E(5) are not contained within the IBA model space, the IBA can reproduce X(5)-like features within a restricted region of parameter space near the critical point, enabling a meaningful comparison between geometric and algebraic approaches. They represent solutions of the Bohr Hamiltonian for specific potentials and should not be interpreted as dynamical symmetries of the IBA.

\begin{figure*}[t]
\centering
\setlength{\unitlength}{0.05\textwidth}
\begin{picture}(10,8.0)
\put(-9.8,-7.5){\includegraphics[width=1.45\textwidth, angle = 0]{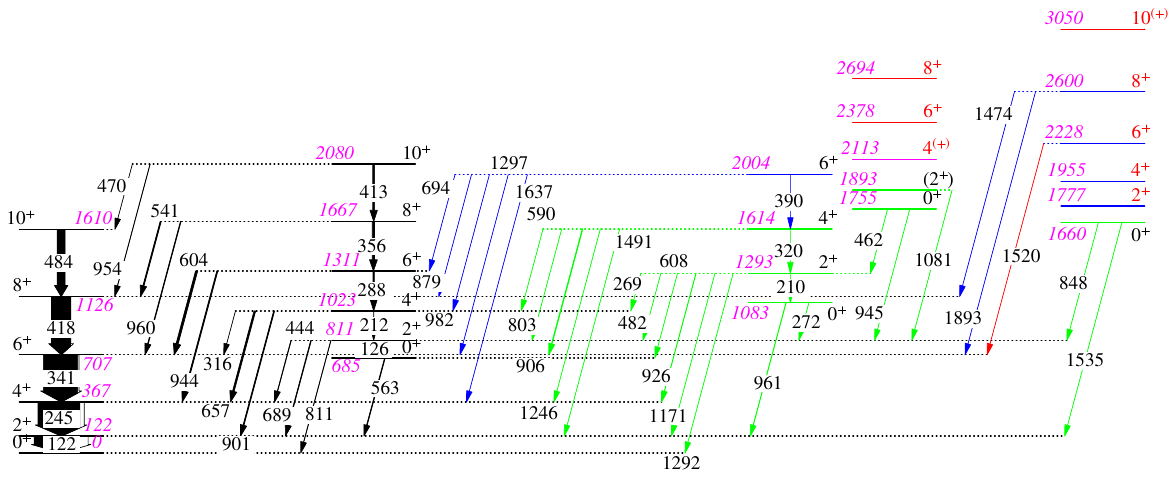}}
\put(-3.0,5.5){\includegraphics[width=0.6\textwidth,angle=0]{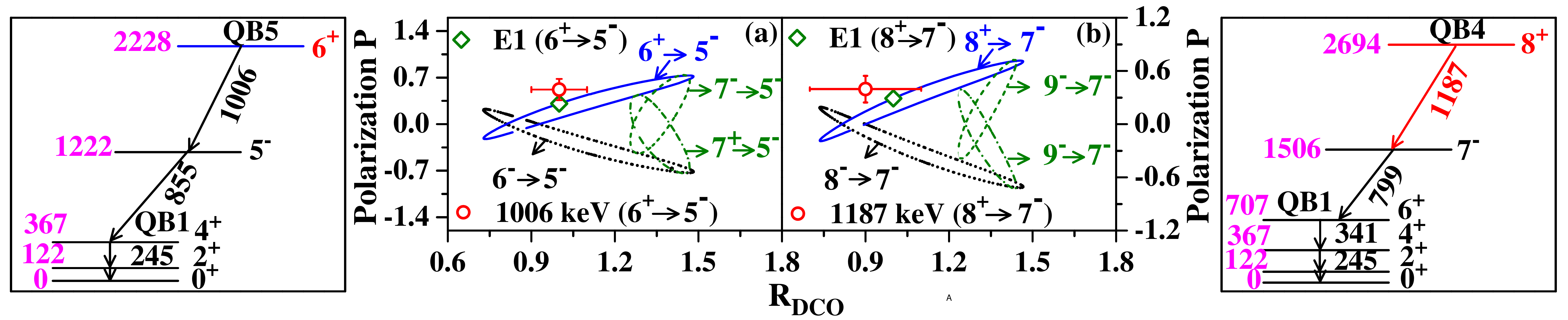}}
\put(-4.80,4.7){QB1}
\put(-5.00,+0.10){\textcolor{black}{\scriptsize{1}}}
\put(-5.00,+0.40){\textcolor{black}{\scriptsize{1}}}
\put(-5.00,+1.00){\textcolor{black}{\scriptsize{1}}}
\put(-5.00,+1.85){\textcolor{black}{\scriptsize{1}}}
\put(-4.95,+4.05){\textcolor{black}{\scriptsize{1}}}
\put(-5.00,+2.85){\textcolor{black}{\scriptsize{1}}}
\put(+0.7,5.4){QB2}
\put(+1.80,+1.80){\textcolor{black}{\scriptsize{2}}}
\put(+1.85,+2.05){\textcolor{black}{\scriptsize{2}}}
\put(+1.75,+2.50){\textcolor{black}{\scriptsize{2}}}
\put(+1.80,+3.25){\textcolor{black}{\scriptsize{2}}}
\put(+1.85,+4.10){\textcolor{black}{\scriptsize{2}}}
\put(+1.90,+5.05){\textcolor{black}{\scriptsize{2}}}
\put(+8.0,5.3){QB3}
\put(+9.20,+2.25){\textcolor{black}{\scriptsize{3}}}
\put(+9.20,+3.25){\textcolor{black}{\scriptsize{3}}}
\put(+9.20,+4.00){\textcolor{black}{\scriptsize{3}}}
\put(+9.15,+4.95){\textcolor{black}{\scriptsize{3}}}
\put(+9.8,7.4){QB4}
\put(+10.90,+4.30){\textcolor{black}{\scriptsize{5}}}
\put(+11.00,+4.70){\textcolor{black}{\scriptsize{5}}}
\put(+11.00,+5.20){\textcolor{red}{\scriptsize{5}}}
\put(+11.00,+5.80){\textcolor{red}{\scriptsize{5}}}
\put(+11.00,+6.60){\textcolor{red}{\scriptsize{4}}}
\put(+13.5,8.0){QB5}
\put(+14.55,+3.65){\textcolor{black}{\scriptsize{4}}}
\put(+14.65,+4.35){\textcolor{red}{\scriptsize{4}}}
\put(+14.65,+4.80){\textcolor{red}{\scriptsize{4}}}
\put(+14.65,+5.50){\textcolor{red}{\scriptsize{4}}}
\put(+14.65,+6.35){\textcolor{red}{\scriptsize{3}}}
\put(+14.75,+7.35){\textcolor{red}{\scriptsize{3}}}
\end{picture}
\caption{\label{levelscheme1} The partial level scheme of $^{152}$Sm obtained in the present work. The level energies (represented by the magenta colored numbers) and the $\gamma$-ray energies are rounded off to the nearest keV. The widths of the arrows are proportional to the intensities of the $\gamma$ -ray transitions. The newly observed levels and transitions are shown in red while the green and blue colored transitions have been first observed in fusion reactions and were seen previously in $\beta$-decay and Coulomb excitations measurements, respectively \cite{bnlnndc1, mjmnd114}. Magenta colored 2113-keV level was previously observed in the $^{153}$Eu(t,$^4$He) reaction \cite{crhirin}. The spin-parities of the levels that are marked in red were not known earlier and have been determined in the present work, as demonstrated in the inset for the (a) 2228~keV and (b) 2694~keV levels of QB5 and QB4, respectively. Here, the indexing applies only to the states observed in this experiment, and there are additional 0$^{+}$, 2$^{+}$, and 4$^{+}$ levels, that were observed previously, in $^{152}$Sm. Similar to those known for QB2 and QB3, we have identified several new parity-changing transitions decaying from the levels of QB4 and QB5 to the negative‐parity states of $^{152}$Sm (two such, 1006 and 1187~keV, are shown in the inset).}
\end{figure*}

The study of quantum shape-phase transitions in atomic nuclei is naturally formulated within the framework of the Interacting Boson Model (IBM), which has successfully described the structural evolution of nuclei in the transitional Nd-Sm-Gd region~\cite{IBM,Scholten1978,Zamfir2004}. In particular, the first-order transition between the spherical $U(5)$ and axially deformed $SU(3)$ dynamical symmetries has attracted considerable attention following the introduction of the X(5) critical-point symmetry by Iachello~\cite{fiachel3}. Among the proposed candidates, the $N=90$ isotones, especially $^{152}$Sm, $^{150}$Nd, $^{154}$Gd, and $^{156}$Dy \cite{Zamfir2004,casten,krucken}, exhibit excitation-energy ratios and electromagnetic transition strengths that closely follow the parameter-free predictions of the X(5) model, supporting their interpretation as nuclei located near the critical point of the $U(5)$--$SU(3)$ phase transition~\cite{casten,krucken}. Particularly noteworthy is $^{152}$Sm, which has emerged as a benchmark realization of X(5)-like behavior, with the observed low-lying excitation spectrum and intraband $B(E2)$ values showing excellent agreement with the theoretical predictions~\cite{casten}. Similar features have been identified in $^{150}$Nd, where both the ground-state ($s=1$) and first excited ($s=2$) bands exhibit the characteristic level structure associated with the zeros of the Bessel functions that arise in the X(5) solution~\cite{krucken,fiachel3}. Nevertheless, significant discrepancies remain. In particular, the measured interband $B(E2)$ transition strengths connecting the $s=2$ and $s=1$ bands in $^{152}$Sm and $^{150}$Nd are systematically weaker than predicted, differing from the X(5) values by approximately a factor of three~\cite{casten,krucken}. Such deviations indicate that finite nuclei do not realize the exact critical-point limit and reflect the influence of finite-size effects, configuration mixing, and departures from the idealized infinite square-well potential in the $\beta$ degree of freedom.

\begin{figure*}[t]
\centering
\setlength{\unitlength}{0.05\textwidth}
\begin{picture}(22,9.8)
\put(+1.5,-0.3){\includegraphics[width=0.85\textwidth, angle = 0]{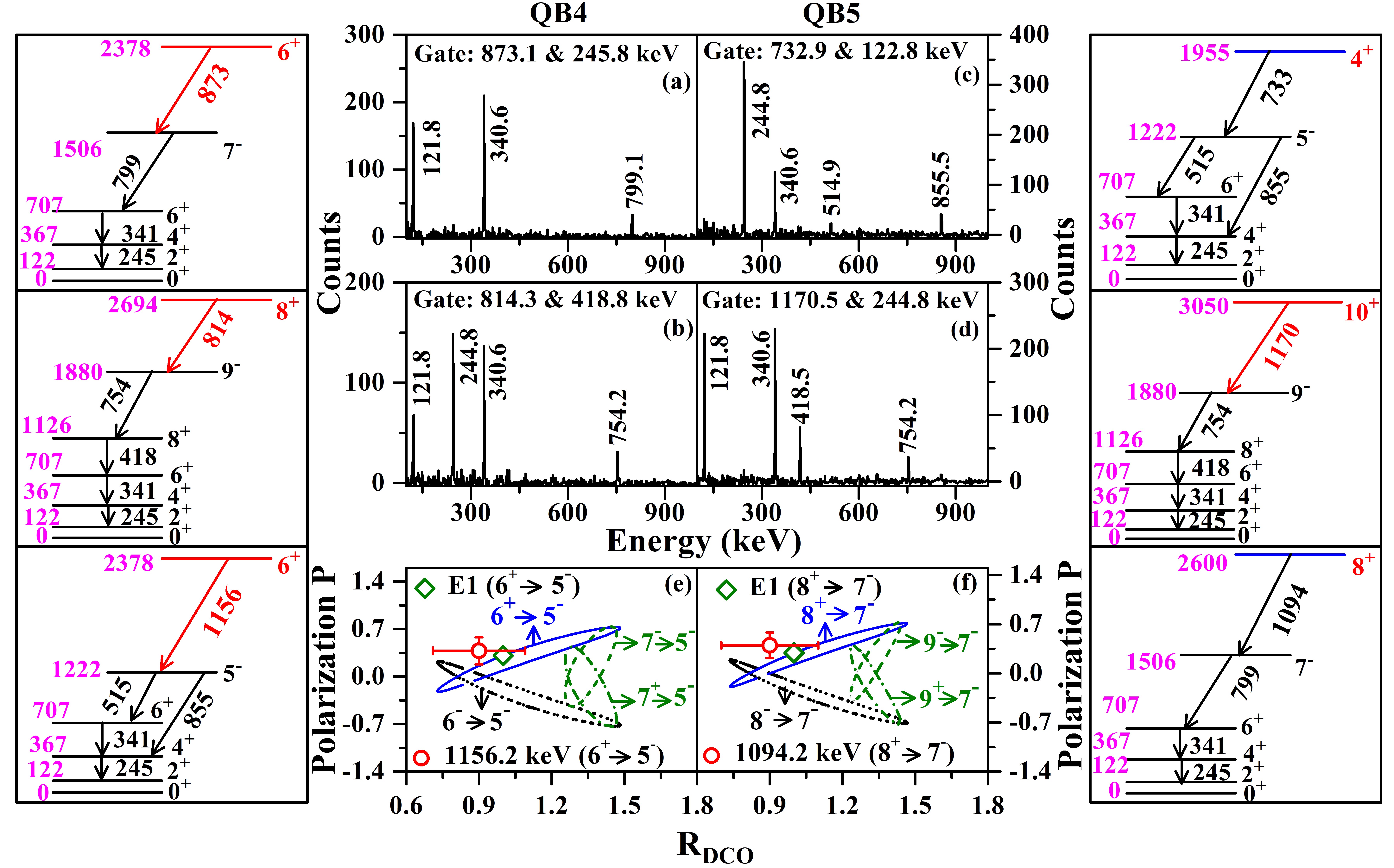}}
\end{picture}
\caption{\label{Fig_04_2} Projections from triple $\gamma$ cube are shown to demonstrate the analysis and observation based on which the new levels were placed in QB4 (left) and QB5 (right) bands. Similar to QB2 and QB3, parity changing $\gamma$ transitions were found that connect these levels to the negative parity states of the octupole band, as also shown in Fig.~\ref{levelscheme1}. The multipolarity of these transitions were found from DCO and Polarization analysis, some of which are displayed in Fig.~\ref{levelscheme1} as well as in this figure. Colour scheme is same as Fig.~\ref{levelscheme1}.}  
\end{figure*}

Similar conclusions have been reached in studies of heavier rare-earth and neutron-rich nuclei, where excitation energies alone often suggest critical-point behavior, while electromagnetic transition strengths reveal a more rotor-like evolution \cite{McCutchan2006}. It is worth noting that most experimental and theoretical investigations of X(5) nuclei have focused primarily on collective structures built upon the first two $0^+$ bandheads. Consequently, the spectroscopic properties of higher-lying collective excitations remain largely unexplored. This limitation highlights the need for more extensive experimental studies extending to higher excitation energies, together with theoretical approaches capable of simultaneously reproducing excitation spectra and electromagnetic transition strengths over an expanded range of collective states. Such investigations are essential for assessing the extent to which finite nuclei realize the X(5) critical-point symmetry and for elucidating the role of finite-size effects and configuration mixing in transitional nuclear systems.


In the present Letter, we report new spectroscopic information on $^{152}$Sm obtained from the $^{150}$Nd($^{4}$He,2n) reaction. The level scheme is extended to include higher-lying band structures built on excited $0^+$ states, providing an expanded experimental basis for examining X(5)-like features in this nucleus. The measured excitation energies, and $B(E2)$ values are compared with IBA calculations performed near the critical region of the U(5)-SU(3) transition. 


\begin{table}[t]
\centering
\scriptsize{}
\caption{\label{table-x51} $\gamma$-energy ($E_\gamma$), branching (Br), mixing ratio ($\delta$), lifetimes, and the corresponding $B(E2)$ transition rates for the states in $^{152}$Sm. Quoted errors of the $B(E2)$ values include the uncertainty due to the $\delta$ (for $E2/M1$ transitions only), Br, and lifetime measurements.}
\begin{tabular}{cccccc}
\hline\hline
\text{$J_{i}$$^{\pi}$ $\rightarrow$ $J_{f}$$^{\pi}$} & \text{$E_\gamma$} & Br   & $\delta$ & $\tau$$^a$ & $B(E2)$ \\

[ $\hbar$ ]                                      & [ keV ]           & [\%]  &          & [ps]   & [W.u.]  \\

\hline\hline

2$^{+}_1$ $\rightarrow$ 0$^{+}_1$  &  121.8(1) & 100     &    & 2024.5(159) & 144.8(11) \\
4$^{+}_1$ $\rightarrow$ 2$^{+}_1$   &  244.8(1) & 100     &    & 83.3(9)     & 208.8(23) \\
6$^{+}_1$ $\rightarrow$ 4$^{+}_1$   &  340.6(2) & 100     &    & 14.9(2)     & 239(3) \\
8$^{+}_1$ $\rightarrow$ 6$^{+}_1$   &  418.5(2) & 100     &    & 4.4(1)      & 294(7) \\
10$^{+}_1$ $\rightarrow$ 8$^{+}_1$  &  483.9(1) & 100     &    & 2.0(2)      & 315(32) \\ 

0$^{+}_2$ $\rightarrow$ 2$^{+}_1$   & 562.9(1)  & 100     &         & 8.8(2)      & 34(3) \\

2$^{+}_2$ $\rightarrow$ 0$^{+}_2$   & 125.7(3)  & 0.38(9)   &         & 10.7(6)     & 94(23) \\ 
2$^{+}_2$ $\rightarrow$ 4$^{+}_1$   & 444.1(2)  & 18.1(6)   &         &             & 16.3(16) \\
2$^{+}_2$ $\rightarrow$ 2$^{+}_1$   & 688.9(1)  & 60.8(8)   & +15.9(14) &     & 5.9(4)   \\
2$^{+}_2$ $\rightarrow$ 0$^{+}_1$   & 810.6(1)  & 20.7(4)   &                 &     & 0.9(1) \\

4$^{+}_2$ $\rightarrow$ 2$^{+}_2$    & 212.4(1)  & 5.2(3)    &         & 12.0(19)    & 145(24) \\
4$^{+}_2$ $\rightarrow$ 6$^{+}_1$    & 316.1(1)  & 3.0(2)    &         &             & 12.8(22)   \\
4$^{+}_2$ $\rightarrow$ 4$^{+}_1$    & 656.6(2)  & 57.6(8)   & +1.7(4) &             & 4.7(7)    \\
4$^{+}_2$ $\rightarrow$ 2$^{+}_1$    & 901.2(1)  & 34.2(6)   &         &             & 0.8(1)     \\
0$^{+}_3$ $\rightarrow$ 2$^{+}_2$    & 272.2(3)  & 14.9(13)  &         & 22(9)    & 73(30) \\
0$^{+}_3$ $\rightarrow$ 2$^{+}_1$    & 961.1(2)  & 75.9(26)  &         &             & 0.7(3) \\
0$^{+}_3$ $\rightarrow$ 1$^{-}_1$    & 119.3(4)  &  9.2(13)  &         &             & \\

2$^{+}_3$ $\rightarrow$ 0$^{+}_3$    & 209.9(2)  & 3.7(3)    &         & $<$23.1     & $>$56.5\\
2$^{+}_3$ $\rightarrow$ 4$^{+}_2$    & 269.4(3)  & 5.8(4)    &         &             & $>$24.9\\
2$^{+}_3$ $\rightarrow$ 2$^{+}_2$    & 482.2(2)  & 6.8(6)    &  +6.6(11)  &             & $>$1.8 \\ 
2$^{+}_3$ $\rightarrow$ 4$^{+}_1$    & 926.3(1)  & 20.5(8)   &         &             & $>$0.2 \\
2$^{+}_3$ $\rightarrow$ 2$^{+}_1$    & 1171.1(3) & 8.2(6)    &         &             & $>$0.03 \\
2$^{+}_3$ $\rightarrow$ 1$^{-}_1$             & 329.6(3)  & 36.0(21)  &         &             & \\
2$^{+}_3$ $\rightarrow$ 3$^{-}_1$             & 251.8(2)  & 23.1(15)  &         &\\
0$^{+}_4$ $\rightarrow$ 2$^{+}_3$    & 462.2(4) & 2.4(3)     &         & $>$0.4     & $<$48.2 \\
0$^{+}_4$ $\rightarrow$ 2$^{+}_2$    & 944.8(6) & 2.1(3)     &         &             & $<$1.2  \\
0$^{+}_4$ $\rightarrow$ 1$^{-}_1$    & 791.7(4) & 95.5(25)   &         &\\
\hline\hline
\end{tabular}
\raggedright
{\footnotesize $^{a}${Level lifetimes are adopted from Refs. \cite{bnlnndc1, mjmnd114}.}}\\
\end{table}

\begin{table}[ht!]
\begin{scriptsize}
\begin{center}
\caption{The details of the decay transitions from the newly found bands along with the DCO ($R_{DCO}$) and Polarization(P) values. The subscript in $I_f^{\pi}$ shows the band (K$^{\pi}$ = 0$^-$ octupole, ground, $\beta$ and $\gamma$ bands) to which the level decays. The $R_{DCO}$ values for all the transitions were determined in a dipole gate except one transition (1473.7) for which quadrupole gate was used.}
\begin{tabular}{ccccccccc}
\hline
\hline
Band&$I_i^{\pi}$&$\rightarrow I_f^{\pi}$&$E_{\gamma}$ & $R_{DCO}$&$(P)$&Multipolarity\\
       &    &                       &   (keV)     &             &         &                 &       \\
\hline

\\
QB4&0$^{+}_5$ & $\rightarrow 2^+_{\gamma}$&462.2(6)&-       &-        & (E2)\\
        &  & $\rightarrow 1^-_{oct}$ & 791.7(4)  &-       &-        & (E1) \\

        &  & $\rightarrow 2^+_{\beta}$ & 944.8(6)    &-       &-        &  (E2)\\
&(2$^{+}_5$)&$\rightarrow 1^-_{oct}$ & 929.4(2) &  -&        -& (E1)\\
          & &$\rightarrow 2^+_{\beta}$&1080.7(6)   &-       &-        & (M1+E2)\\
&4$^{(+)}_5$  &$\rightarrow 3^-_{oct}$  & 1071.9(2)&  1.0(3) &-        &E1\\
          & &$\rightarrow 4^+_{gsb}$  & 1746.3(3)&-       &-        &(M1+E2) \\
         &  &$\rightarrow 2^+_{gsb}$  & 1990.9(3)&       -&        -&(E2)\\
&6$^+_5$      &$\rightarrow 7^-_{oct}$  & 873.1(2) &       -&        -& E1\\
          & &$\rightarrow 5^-_{oct}$  & 1156.2(1)& 0.9(2) & 0.38(19) & E1\\
&8$^+_4$      &$\rightarrow 9^-_{oct}$  & 814.3(1) & 1.0(3) &        -& E1\\
          & &$\rightarrow 7^-_{oct}$  & 1187.5(1)& 0.9(2) & 0.40(15) & E1\\

QB5&0$^+_4$ & $\rightarrow 1^-_{oct}$   & 696.1(3)  & -      & -       & (E1) \\
     & & $\rightarrow 2^+_{\beta}$ &  847.5(2)  &  -      & -       & (E2) \\
    &  & $\rightarrow 2^+_{gsb}$   & 1535.3(6) & -      &-        & (E2) \\
&2$^+_4$ & $\rightarrow 3^-_{oct}$   & 735.8(1)   & 0.9(2) & 0.12(8) &  E1  \\
     && $\rightarrow 1^-_{oct}$   & 813.4(1)    & 0.8(3) & 0.39(33)  &  E1 \\
&4$^+_4$ & $\rightarrow 5^-_{oct}$   & 732.9(1)    & 1.1(2) & 0.39(19)  &  E1 \\
     & & $\rightarrow 3^-_{oct}$   & 913.4(1)    & 0.9(2) & 0.61(12) &  E1 \\
&6$^+_4$ & $\rightarrow 7^-_{oct}$   & 722.1(1)    & 1.2(2) & 0.41(23)  &  E1 \\
     & & $\rightarrow 5^-_{oct}$   & 1006.1(1)   & 1.0(1) & 0.52(16) &  E1 \\
     & & $\rightarrow 6^+_{gsb}$   & 1519.9(6)   &       -&        -& (M1+E2)\\
&8$^+_3$ & $\rightarrow 9^-_{oct}$   &  719.6(1)   & 1.1(1) & 0.34(9) &  E1 \\
    &  & $\rightarrow 7^-_{oct}$   & 1094.2(1)   & 0.9(2) & 0.40(18) &  E1\\
    &  & $\rightarrow 8^+_{gsb}$   & 1473.7(2)   & 0.9(3) & - & (M1+E2) \\
    &  & $\rightarrow 6^+_{gsb}$   & 1893.3(5)   &       -&       - & (E2) \\
&10$^{(+)}_3$& $\rightarrow 9^-_{oct}$ & 1170.5(1) & 1.1(2) &-        &  E1 \\ 

\hline
\hline
\end{tabular}
\label{tabdco22}
\end{center}
\end{scriptsize}
\end{table}

Excited states of $^{152}$Sm were populated using $^{150}$Nd($^{4}$He,2n) reaction at E$_{lab}$ = 26 MeV and a 10 mg/cm$^2$ $^{150}$Nd (99\% enriched) target.  An efficient array of twelve Compton-suppressed HPGe clover detectors (3 nos at 40$^{\circ}$ and 125$^{\circ}$ each, 6 nos at 90$^{\circ}$ with respect to the beam axis), was employed for $\gamma$-ray detection. Of the total number of single fold events acquired during the experiment, around 93\%, amounting to 39.7 billion, was from the de-excitation of $^{152}$Sm. The data were sorted into different symmetric and angle dependent $\gamma$ - $\gamma$ matrices, as well as $\gamma$ - $\gamma$ - $\gamma$ cube using the IUCPIX \cite{iucpix} code and analyzed using the INGASORT \cite{ingasort} and the RADWARE \cite{radford1, radford2} packages. The multipolarities and the electromagnetic characters of the observed gamma-ray transitions were determined from the measurements of the ratio for Directional Correlation from Oriented state, DCO ratio ($R_{\textrm{DCO}}$) \cite{kramer, kabadi},  and the linear polarization ($ P $) \cite{staro2, droste2, deng2, jones2}. In the present investigation, the detectors at 125$^{\circ}$ and 90$^{\circ}$ with respect to the beam direction were used to evaluate the $R_{DCO}$ values which have been compared with the theoretical DCO ratios \cite{kramer, kabadi} for multipolarity assignments of the $\gamma$ -ray transitions. For the $P$ measurements, the detectors placed at angles 90$^{\circ}$ were used. The  details on experiment and analysis procedures are already reported in Ref.~\cite{shefali_PRC}.

Measured $R_{DCO}$ and $P$ values of $\gamma$-ray transitions have been used for the spin parity assignments. For each level, all possible multipolarities (E1, M1, and E2) were considered, and the adopted assignments correspond to the only values consistent with the observed transition strengths and selection rules. 
The  $R_{DCO}$ and $P$ data have also been used to extract the mixing ratio ($\delta$) of the mixed $E2/M1$ transitions by comparing their theoretical values. The intensities of the $\gamma$-ray transitions have been determined from the gated projections of the symmetric $E_{\gamma}$ - $E_{\gamma}$ matrix and normalized to the intensity of the 121.8-keV (2$^{+}$ $\rightarrow$ 0$^{+}$) transition. 
The branching ratios and $\delta$ (for $E2/M1$ transitions only) values obtained from the present measurement when combined with the previously reported level lifetimes \cite{bnlnndc1, mjmnd114, jhhamilt, rfcast22}  produces an accurate determination of the B(E2) transition probabilities, tabulated in Table \ref{table-x51}. The $\gamma$-ray transition energy ($E_{\gamma}$), $R_{\textrm{DCO}}$, $ P $ and the spin-parity of the newly established level structure of $^{152}$Sm are recorded in Table \ref{tabdco22}, and are in agreement with the literature values \cite{bnlnndc1}, wherever known. Some of the representative results from analyses of $R_{DCO}$ and $P$  have also been displayed in Fig.~\ref{levelscheme1} and Fig.~\ref{Fig_04_2}.

The positive parity states of $^{152}$Sm obtained from the present work (Fig.~\ref{levelscheme1}) are in agreement with the previously reported structures ~\cite{bnlnndc1, mjmnd114, jhhamilt, rfcast22, crhirin, kulp2005, kulp2008}. The previously known levels along with the newly found ones have been extended to five band structures QB1, QB2, QB3, QB4 and QB5 above the 0.0, 684.8, 1082.9, 1755.3, 1659.7-keV 0$^+$ states, respectively, through the observation of new transitions, as shown in Fig.~\ref{levelscheme1} using the coincidence analysis of triple $\gamma$ cube displayed in Fig.~\ref{Fig_04_2}. The band structures QB3, QB4 and QB5; and the interconnecting transitions to QB1, QB2 and QB3 have been observed for the first time in fusion evaporation reaction. Among these, the transitions decaying from the 0$^+_3$, 1082.9-keV and 2$^+_3$, 1292.9-keV states of QB3 and 2$^+_5$, 1893.0-keV level of QB4 were observed in the $\beta$-decay of $^{152}$Pm~\cite{hmach11} whereas 0$^+_3$, 1082.9-keV; 2$^+_3$, 1292.9-keV and 4$^+_3$, 1613.7-keV states of QB3 were observed in the $\epsilon$-decay of $^{152}$Eu \cite{bnlnndc1, mjmnd114}.

It is to be noted that the QB1 - QB5 bands are placed on the first five 0$^+$ levels in $^{152}$Sm, confirmed in the present work.  Several other previously reported $0^+$, $2^+$, $4^+$, $6^+$, $8^+$ and $10^+$ states within the same energy range have not been included in Fig.~\ref{levelscheme1}, not to obscure the clarity of the present observation. These states are indicated in table~\ref{tab1}. Most of these levels are candidates of other band structure except 1226~keV and 1736~keV states, presence of which could not be confirmed from the present data.



The newly identified levels in bands QB4 and QB5 show no observed intra-band E2 transitions, unlike bands QB2 and QB3. Their energies, however, exhibit a parabolic progression nearly identical to those bands. Our sensitivity limit of $\sim$ 0.01\% on relative intensity implies any intra-band E2 branch would lie below this threshold. Crucially, the $\gamma$-transition expected from the 0$^+_4$ (1659.7 keV) state of QB5 to the 2$^+_3$ (1292.9 keV) state of QB3 remains unobserved; given that a sizable branch would be anticipated if QB5 were part of the X(5) structure, we therefore do not include QB5 (R$_{4/2}$ = 2.52)  in the X(5)-like band sequence discussed below.


\begin{figure*}[t]
\centering
\setlength{\unitlength}{0.05\textwidth}
\begin{picture}(10,8.6)
\put(+1.6,-3.4){\includegraphics[width=0.82\textwidth, angle = 0]{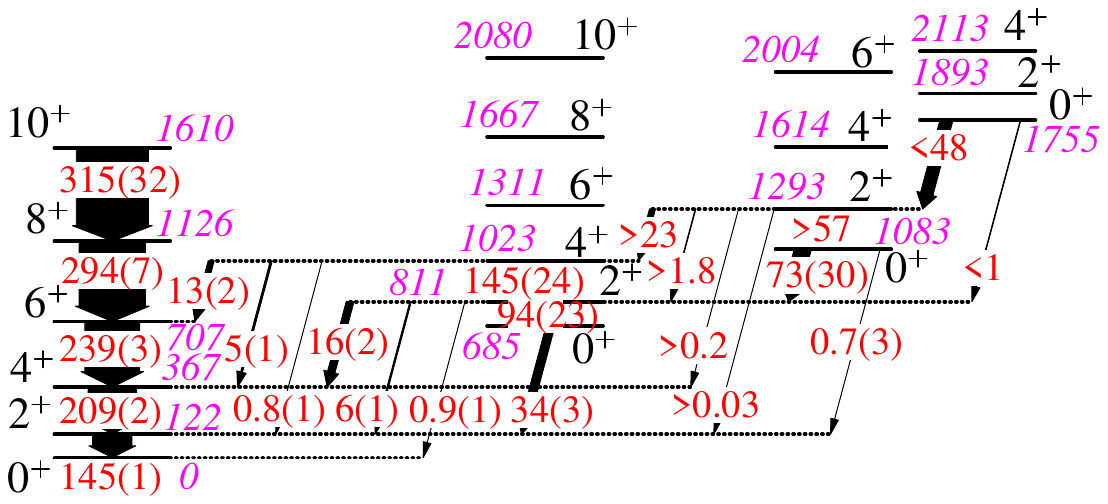}}
\put(-7.7,-1.2){\includegraphics[width=0.75\textwidth, angle = 0]{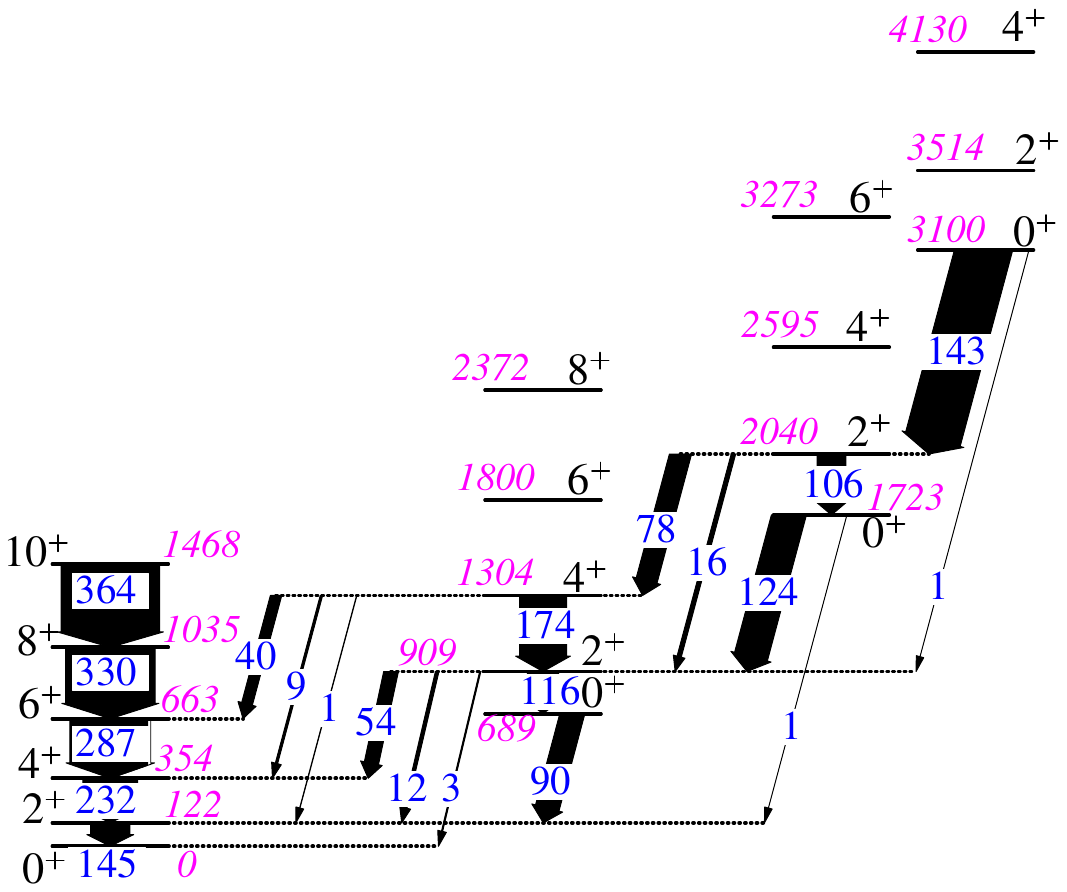}}
\put(-1.8,0.6){(a)}
\put(+8.2,0.6){(b)}
\put(+5.20,+4.65){\textcolor{red}{\scriptsize{$s$ = 1}}}
\put(+5.00,+5.00){\textcolor{red}{\scriptsize{R$_{4/2}$ = 3.01}}}
\put(9.30,+5.45){\textcolor{red}{\scriptsize{$s$ = 2}}}
\put(9.00,+5.75){\textcolor{red}{\scriptsize{R$_{4/2}$ = 2.68}}}
\put(+12.10,+5.35){\textcolor{red}{\scriptsize{$s$ = 3}}}
\put(+11.70,+5.70){\textcolor{red}{\scriptsize{R$_{4/2}$ = 2.53}}}
\put(+13.80,+5.60){\textcolor{red}{\scriptsize{$s$ = 4}}}
\put(+13.50,+5.95){\textcolor{red}{\scriptsize{R$_{4/2}$ = 2.59}}}

\put(-4.50,+3.95){\textcolor{blue}{\scriptsize{$s$ = 1}}}
\put(-4.7,+4.30){\textcolor{blue}{\scriptsize{R$_{4/2}$ = 2.90}}}
\put(-0.70,+5.65){\textcolor{blue}{\scriptsize{$s$ = 2}}}
\put(-1.00,+6.00){\textcolor{blue}{\scriptsize{R$_{4/2}$ = 2.80}}}
\put(+1.80,+7.20){\textcolor{blue}{\scriptsize{$s$ = 3}}}
\put(+1.60,+7.55){\textcolor{blue}{\scriptsize{R$_{4/2}$ = 2.75}}}
\put(+3.35,+8.70){\textcolor{blue}{\scriptsize{$s$ = 4}}}
\put(+3.10,+9.05){\textcolor{blue}{\scriptsize{R$_{4/2}$ = 2.49}}}
\end{picture}
\vspace{-0.7cm}
\caption{\label{level-iba3} The level scheme of $^{152}$Sm obtained from (a)  the X(5) critical-point solution (for $s = 1, 2, 3,$ and $4$ family of bands)  and (b) the present experiment. The level energies, both for the experimental and calculated scheme, are rounded off to the nearest keV (represented by the magenta-colored numbers). The label (blue in (a) and colored red in (b)) associated with the transitions represents the B(E2) values in W.u. and normalized with the measured value of B$(E2: 2^+_{1} \rightarrow 0^+_{1})$ = 144.8 W.u.. B$(E2)$ values greater than one in the unit of W.u. are rounded off to the nearest integer whereas for the value less than one, up to the first decimal place has been considered except for the 2$^+_3$ $\rightarrow$ 2$^+_1$ transition, for which B$(E2: 2^+_{3} \rightarrow 2^+_{1})$ is $>$ 0.03 W.u..}  
\end{figure*}

\begin{table}[t]
\begin{scriptsize}
\centering
\caption{Levels excluded from the proposed QB4 and QB5 bands and the reasons for their exclusion. Assignment to different band structure is as per ENSDF.}
\begin{tabular}{ccc}
\hline
 Known $J^{\pi}$  & Energy (keV) & Remarks \\ 
 values from&&\\
  ENSDF&&\\
\hline
$0^+$ & 1736 & Not confirmed in the present work. \\ 

$2^+$ & 1086 & Member of the $K^{\pi} = 2^+$ $\gamma$-vibrational band \\ 
$(2^+)$&1226& Not confirmed in the present work. \\
$2^+$&1769&  Member of the $K^{\pi} = 2^+$ band\\

$4^+$ & 1372 & Member of the $K^{\pi} = 2^+$ $\gamma$-vibrational band \\
$4^+$ &1757& Member of the $K^{\pi} = 4^+$ band \\
$4^+$ &2052&Member of the $K^{\pi} = 2^+$ band \\
$6^+$ & 1728 & Member of the $K^{\pi} = 2^+$ $\gamma$-vibrational band \\
$6^+$&2040& Member of the $K^{\pi} = 4^+$ band\\
$8^+$ & 2140 & Member of the $K^{\pi} = 2^+$ $\gamma$-vibrational band \\
$8^+$&2392& Member of the $K^{\pi} = 4^+$  band\\
$8^+$&2459&Band head of the $K^{\pi} = 8^+$  band\\
$10^+$ & 2662 & Member of the $K^{\pi} = 2^+$ $\gamma$-vibrational band \\
$(10^+)$&2810& Member of the $K^{\pi} = 4^+$  band\\
$10^+$&2905& Member of the $K^{\pi} = 8^+$  band\\

\hline
\end{tabular}
\label{tab1}
\end{scriptsize}
\end{table}

The observed level structure of $^{152}$Sm (Fig.~\ref{levelscheme1}) exhibits distinctive signatures of the X(5) critical-point symmetry, which describes nuclei at the first-order phase transition between the spherical $U(5)$ and axially deformed $SU(3)$ limits. The positive-parity bands built on the ground $0^+_1$ and low-lying $0^+_2$ states display excitation-energy systematics and rotational sequences that closely follow the characteristic X(5) pattern (Fig. \ref{level-iba3}(a)). In particular, the measured excitation-energy ratios $R_{4/2}=E(4_1^+)/E(2_1^+)=3.01$ and $E(0_2^+)/E(2_1^+)=5.62$ are in excellent agreement with the corresponding X(5) predictions of 2.91 and 5.67, respectively. The yrast-band energies and intraband $B(E2)$ transition strengths are likewise reproduced reasonably well within the parameter-free X(5) framework.

The new band structures added to the experimental level scheme shown in Fig.~\ref{levelscheme1} reveals extended collective structures built upon the $0_3^+$, 1082.9 keV  and $0_4^+$, 1755.3 keV states, thereby providing an opportunity to examine the applicability of the X(5) description beyond the lowest-lying bands. It follows from the excitation scheme of $^{152}$Sm, Fig. \ref{level-iba3}(b), the 0$^{+}_3$ and 0$^{+}_4$ states at 1082.9 keV and 1755.3 keV have a large B(E2) value to the 2$^{+}_{2}$ and 2$^{+}_{2}$ states and have thus been identified to be bandhead of the s = 3 and s = 4 family of levels. Measured values $\frac{B(E2: 0^+_{3} \rightarrow 2^+_{2})}{B(E2: 0^+_{3} \rightarrow 2^+_{1})}$ = 104(5) and $\frac{B(E2: 0^+_{4} \rightarrow 2^+_{3})}{B(E2: 0^+_{4} \rightarrow 2^+_{2})}$ = 40.2(76) are in satisfactory agreement with the X(5) predictions 124, and 143, respectively. These observations further corroborate the assignment of the states of 1082.9 keV and 1755.3 keV as 0$^+_{3}$, s = 3 and 0$^+_{4}$, s = 4 levels of the critical point symmetry X(5), respectively.

\begin{figure}[t]
\centering
\setlength{\unitlength}{0.05\textwidth}
\begin{picture}(10,7.8)
\put(+0.2,-7.1){\includegraphics[width=0.75\textwidth, angle = 0]{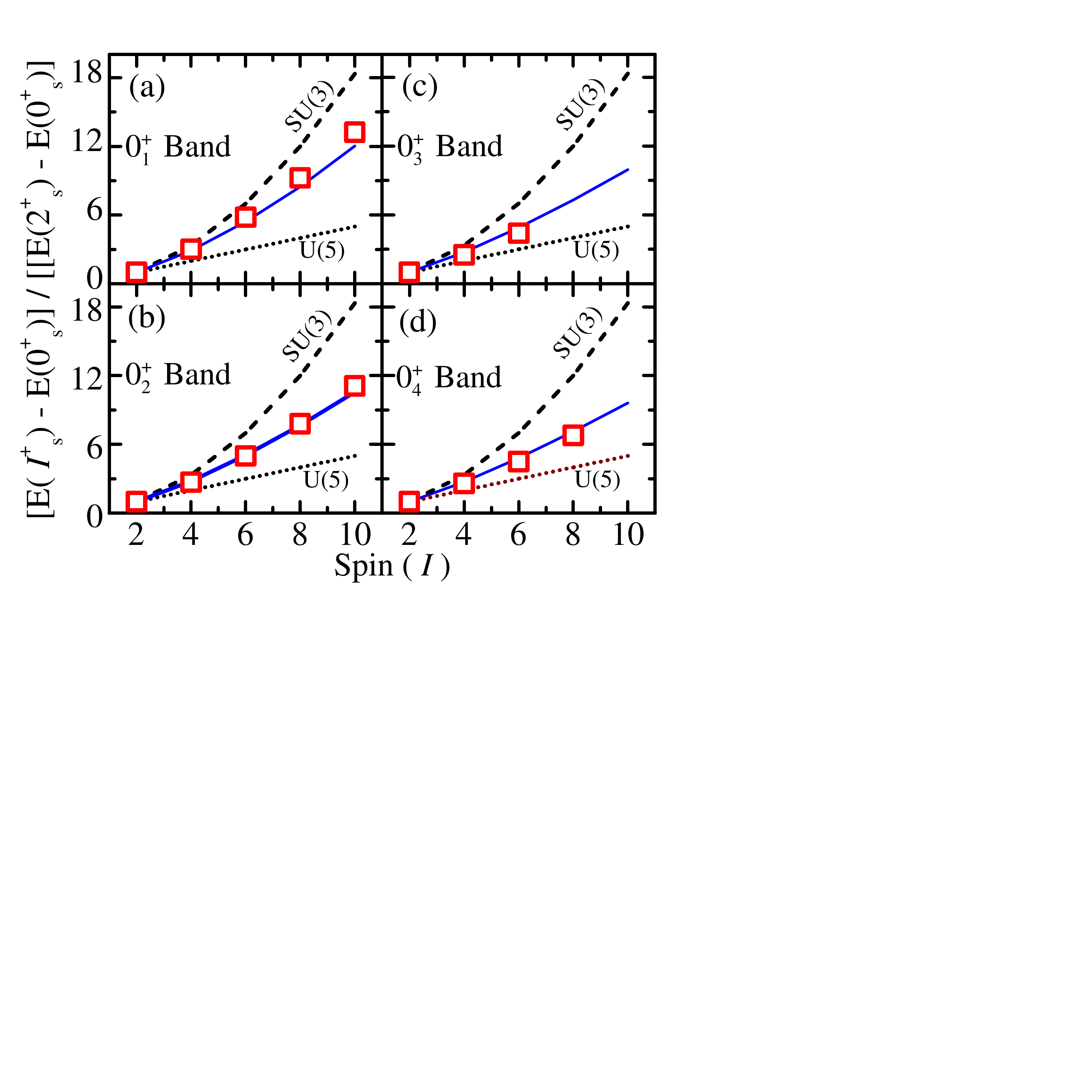}}
\end{picture}
\caption{\label{level-iba2} Comparison of the experimental energy ratios (red squares) with those obtained from the X(5) model calculations (solid blue lines). The limiting values corresponding to the axial rotor SU(3) and spherical vibrator U(5) symmetries are indicated by black short-dashed and dotted lines, respectively.}
\end{figure}

As illustrated in Fig.~\ref{level-iba2} and Table \ref{tab4}, the relative energy ratios of the observed band structures follow the overall trend predicted by the X(5). The evolution of the experimental $R_{4/2}$ values with increasing band-head excitation energy also exhibits the characteristic behavior expected for X(5)-like nuclei. Fig. \ref{level-iba3} illustrates good agreement of the (experimental) level energies and the associated B(E2) values in $^{152}$Sm, with the predictions of the X(5) symmetry. While the calculated level energies therefrom exhibit reasonable overlap with the experimental ones, the in-band B(E2) values  from the (present) measurement are in excellent compliance with those from the calculations normalized with B$(E2: 2^+_{1} \rightarrow 0^+_{1})$ = 148.8 W.u.. These observations collectively indicate the existence of X(5) symmetry extended to s = 4 family in $^{152}$Sm.


Despite the overall agreement with the X(5) predictions, systematic deviations become evident for the higher-lying collective excitations. While the relative band structures are reproduced qualitatively, the excitation energies of the higher $0^+$ states are overestimated by approximately 300-400~keV, and discrepancies persist in several interband transition strengths (Fig. \ref{level-iba3} and Table \ref{tab4}). These observations indicate that the idealized X(5) framework captures the dominant collective features of $^{152}$Sm but does not fully account for the detailed structure of the excited configurations. The deviations point to the influence of finite-size effects, configuration mixing, and a softer $\beta$ potential than the infinite square-well form assumed in the original X(5) formulation. Consequently, $^{152}$Sm should be regarded not as an exact realization of the X(5) critical-point symmetry, but rather as a nucleus located in its immediate vicinity within the $U(5)$--$SU(3)$ shape-phase transitional region.

\begin{table}[t]
\begin{center}
\caption{Comparison of the ratios of the experimental and X(5) model predicted transition probabilities and excitation energies of the levels within and out of different $s$ families. The ratio $B_{J_{in} \rightarrow J_{fm}} = \frac{B(E2: J_i^+({s_n}) \rightarrow J_f^+({s_m}))}{B(E2: 2^+_{1}\rightarrow 0^+_{1})}$, where m,n can vary from 1 to 5 and $n \geq m$.}
\begin{scriptsize}
\begin{tabular}{cccccccc}
\hline\hline
Ratio & Expt. & X(5) & & Ratio & Expt. & X(5) \\
\hline
B$_{41 \rightarrow 21}$ & 1.44(2) & 1.60 & &  B$_{42 \rightarrow 22}$ & 1.00(17)& 1.20 \\
B$_{61 \rightarrow 41}$ & 1.65(3) & 1.98 & &  B$_{42 \rightarrow 61}$ & 0.09(2) & 0.28 \\
B$_{81 \rightarrow 61}$ & 2.03(5) & 2.28 & &  B$_{42 \rightarrow 41}$ & 0.03(1) & 0.06 \\
B$_{81 \rightarrow 61}$ & 2.18(22)& 2.51 & &  B$_{42 \rightarrow 21}$ & 0.006(1)& 0.009 \\

B$_{02 \rightarrow 21}$ & 0.24(2) & 0.62 & &  B$_{03 \rightarrow 22}$ & 0.50(21)& 0.86 & \\

B$_{22 \rightarrow 02}$ & 0.65(16)& 0.80 & &  B$_{23 \rightarrow 03}$ & $>$0.39 & 0.24 &\\
B$_{22 \rightarrow 41}$ & 0.11(1) & 0.37 & &  B$_{23 \rightarrow 42}$ & $>$0.17 & 0.11 &\\
B$_{22 \rightarrow 21}$ & 0.04(1) & 0.08 & &  B$_{04 \rightarrow 23}$ & $<$0.33 & 0.99 & \\

$\frac{E(2^+_{2})}{E(2^+_{1})}$& 6.65 & 7.45 & & $\frac{E(0^+_{3})}{E(2^+_{1})}$& 8.88 & 14.12 & \\
$\frac{E(0^+_{4})}{E(2^+_{1})}$& 14.39& 25.41 & \\

\hline\hline
\end{tabular}
\end{scriptsize}
\label{tab4}
\end{center}
\end{table}

The discrepancies observed between the experimental data of $^{152}$Sm and the ideal X(5) predictions are consistent with finite-size effects expected in realistic nuclei. IBM studies have shown that the first-order $U(5)$--$SU(3)$ phase transition is significantly broadened for finite boson numbers, resulting in a critical region rather than a unique critical point~\cite{Rowe2005}. Furthermore, candidate X(5) nuclei such as $^{152}$Sm are predicted to lie close to, but not exactly at, the critical point in the IBM symmetry triangle~\cite{Zamfir2004}. Since critical-point symmetries represent limiting large-$N_B$ solutions, finite boson-number corrections are expected to affect both excitation energies and electromagnetic transition strengths~\cite{Arias2003}. Consequently, the observed compression of the excited $0^+$ structures and the systematic reduction of interband $B(E2)$ strengths relative to the X(5) predictions may be interpreted as manifestations of finite-size effects, configuration mixing, and deviations from the idealized infinite square-well potential in the $\beta$ degree of freedom.

An additional source of the discrepancies between the experimental data of $^{152}$Sm and the original X(5) predictions may be traced to the assumed shape of the collective potential in the $\beta$ degree of freedom. The X(5) solution employs an infinite square-well potential, corresponding to a completely flat potential with infinitely rigid walls~\cite{fiachel3}. Although this approximation captures the essential features of a nucleus near the first-order phase-transition point, it represents an extreme limit that is unlikely to be realized in finite nuclei. Caprio demonstrated that replacing the infinite wall by a finite-slope potential leads to a substantial compression of the excited $\beta$-band energies while leaving the ground-state band largely unaffected~\cite{Caprio2004}. In particular, the experimentally observed spacing within the $0_2^+$ band of $^{152}$Sm is significantly smaller than predicted by X(5), suggesting a softer effective $\beta$ potential than the idealized infinite square well. Similar conclusions were reached in studies of the X(5)-$\beta^2$, X(5)-$\beta^4$, and related models, where smoother potentials were shown to provide an improved description of excited $0^+$ structures and $\beta$-band spacings while retaining the principal signatures of critical-point behavior~\cite{Bonatsos2004,Bonatsos2007}. These results indicate that the deviations observed in $^{152}$Sm may reflect not only finite-size effects but also the sensitivity of critical-point observables to the detailed shape and stiffness of the underlying collective potential.

In summary, the positive-parity structure of $^{152}$Sm has been investigated through the $^{150}$Nd($\alpha$,2n)$^{152}$Sm reaction, leading to the observation of several new levels and $\gamma$-ray transitions. The level scheme has been extended to five collective band structures built on the lowest 0$^+$ states, and spin-parity assignments have been established from directional-correlation and polarization measurements. The observed excitation energies and B(E2) transition strengths support the interpretation of the bands built on the 0$^+_3$ and 0$^+_4$ states as excited collective structures exhibiting X(5)-like characteristics, extending the known X(5)-like sequence beyond the well-established s=1 and s=2 families. While the overall energy and transition-rate systematics are broadly consistent with the X(5) critical-point description, systematic deviations are observed for higher-lying excitations, reflecting the influence of finite-size effects, configuration mixing, and departures from the idealized infinite square-well potential. These results provide new experimental constraints on the manifestation of critical-point behavior in finite nuclei.

\section*{Acknowledgements}

Authors gratefully acknowledge the efforts of the K-130 cyclotron operation team for providing good quality $^4$He beam. The authors also acknowledge the efforts of those who assisted in setting up and maintaining the array.

\bibliographystyle{unsrtnat} 
\bibliography{Sm152_PLB_F}

\end{document}